\documentclass[12pt]{article}
\usepackage{amssymb,array,longtable,cite}
\def\p#1{\phi_{ #1}}
\def\ps#1{\phi^*_{ #1}}

\newcommand{\drawsquare}[2]{\hbox{%
\rule{#2pt}{#1pt}\hskip-#2pt
\rule{#1pt}{#2pt}\hskip-#1pt
\rule[#1pt]{#1pt}{#2pt}}\rule[#1pt]{#2pt}{#2pt}\hskip-#2pt
\rule{#2pt}{#1pt}}
\def\Bbb{\@subst@obsolete{amsfonts}\Bbb\mathbb}

\newcommand{\Yfund}{\raisebox{-.5pt}{\drawsquare{6.5}{0.4}}}
\newcommand{\Ysymm}{\raisebox{-.5pt}{\drawsquare{6.5}{0.4}}\hskip-0.4pt%
        \raisebox{-.5pt}{\drawsquare{6.5}{0.4}}}
\newcommand{\Yasymm}{\raisebox{-3.5pt}{\drawsquare{6.5}{0.4}}\hskip-6.9pt%
        \raisebox{3pt}{\drawsquare{6.5}{0.4}}}
\newcommand{\Ythreea}{\raisebox{-3.5pt}{\drawsquare{6.5}{0.4}}\hskip-6.9pt%
        \raisebox{3pt}{\drawsquare{6.5}{0.4}}\hskip-6.9pt
        \raisebox{9.5pt}{\drawsquare{6.5}{0.4}}}

\newcommand{\jref}[4]{{\it #1} {\bf #2}, #3 (#4)}

\newcommand{\MPLA}[3]{\jref{Mod.\ Phys.\ Lett.}{A#1}{#2}{#3}}

\newcommand{\NPB}[3]{\jref{Nucl.\ Phys.}{B#1}{#2}{#3}}

\newcommand{\PLB}[3]{\jref{Phys.\ Lett.}{#1B}{#2}{#3}}
\newcommand{\PR}[3]{\jref{Phys.\ Rep.}{#1}{#2}{#3}}
\newcommand{\PRD}[3]{\jref{Phys.\ Rev.}{D#1}{#2}{#3}}

\newcommand{\PRL}[3]{\jref{Phys.\ Rev.\ Lett.}{#1}{#2}{#3}}

\renewcommand{\theequation}{\thesection.\arabic{equation}}
\makeatletter
\def\vereq#1#2{\lower3pt\vbox{\baselineskip1.5pt \lineskip1.5pt
\ialign{$\m@th#1\hfill##\hfil$\crcr#2\crcr\sim\crcr}}}
\makeatother

\begin{document}

\begin{titlepage}
\begin{center}
hep-th/9803087     \hfill   UCSD/PTH 98-09 \\

\vskip 1.3in
 
{\Large \bf Supersymmetric gauge theories with a free algebra \\ of invariants}

\vskip 0.25in

{\bf Gustavo Dotti,\footnote{Address after April 1, 1998: FaMAF,
Universidad Nacional de C\'ordoba, (5000) C\'ordoba, Argentina.
e-mail: {\tt gdotti@fis.uncor.edu}} Aneesh V.~Manohar, and Witold Skiba}

\vskip 0.15in

{\em Department of Physics, University of California at San Diego, \\
9500 Gilman Drive, La Jolla, CA 92093}

\vskip 0.1in

\end{center}

\vskip .5in

\begin{abstract}
We study the low-energy dynamics of all ${\cal N}=1$ supersymmetric gauge
theories whose basic gauge invariant fields are unconstrained. This set
includes all theories whose matter Dynkin index is less than the index of
the adjoint representation. We study the dynamically generated superpotential
in these theories, and show that there is a $W=0$ branch if and only if
anomaly matching is satisfied at the origin. An interesting example studied
in detail  is $SO(13)$ with a spinor, a theory with a dynamically generated
$W$ and no anomaly matching at the origin. It flows via the Higgs mechanism
to $SU(6)$ with a three-index antisymmetric tensor, a theory with a $W=0$
branch and anomaly matching at the origin.

\end{abstract}

\end{titlepage}

\newpage

\section{Introduction}
\setcounter{equation}{0}
The first step in studying the low-energy dynamics of supersymmetric
gauge theories~\cite{ADS,CERN,Seiberg} is determining the structure of
the moduli space. The moduli space can always
be described in terms of the expectation value of composite fields which are
gauge invariant polynomials constructed out of the microscopic chiral matter
superfields~\cite{ProcSchwarz,BDFS,GA-NPB}. In general,
the gauge invariant polynomials are subject to constraints. One can always
choose a basic set of linearly independent composite fields such that all gauge
invariant composites can be written as polynomials in the basic set.
All constraints among gauge invariant composites are then reduced to nonlinear
constraints among the basic set of invariants.
Given the description of the classical moduli space, one can use symmetry
arguments, the semi-classical limit, and various deformations of the theory
to learn about the quantum behavior.
Deformations such as adding mass terms or breaking the gauge group by the
Higgs mechanism are particularly useful, and one can often use them to relate
one theory to another theory whose low-energy description is already known.

In this article we examine all asymptotically-free theories with a free algebra
of invariants. We consider theories with simple gauge groups and no tree-level
superpotentials. By a free algebra of invariants we mean that the full
classical moduli space is described in terms of independent gauge invariants
which are not subject to any constraints. In a previous paper~\cite{GA-PRL} a
class of free-algebra theories was studied in which flavor anomalies are
saturated by the basic gauge invariants. It was shown that, in most cases,
matching of anomalies together with the requirement of a free algebra  implied
the existence of a branch of the theory with no dynamically generated
superpotential. In this paper, we extend the results of
Ref.~\cite{GA-PRL} to all theories with a free algebra of invariants.

We expect different explanations for the lack of anomaly matching between the
microscopic theory and the composites parameterizing the moduli space depending
on the relative value of the Dynkin index of the matter fields, $\mu$, and the
Dynkin index of the adjoint, $\mu_{adj}$. When $\mu < \mu_{adj}$, a dynamically
generated superpotential is allowed by symmetries. Such a superpotential lifts
the classical moduli space leaving no stable vacuum state. We explicitly check
that in all cases where anomalies do not match a superpotential is indeed
dynamically  generated. Maximally breaking the original gauge group often
leaves a pure Yang-Mills theory with gauge group $G_P$,  and gaugino
condensation in this minimal unbroken subgroup generates a dynamical
superpotential. When the gauge group is completely broken, one can check that
instanton contributions are responsible for generating a dynamical
superpotential. There is an interesting difference between those theories with
$\mu < \mu_{adj}$ where anomalies do match at the origin, and those where they
do not. In theories where the anomalies match at the origin,the minimal
unbroken gauge group $G_P$ is a product group, with identical factors (e.g.
$SU(3) \times SU(3)$). The dynamical superpotential is generated by gaugino
condensation in the components of $G_P$, leading to a multi-branched theory. For
at least one branch of the theory, the superpotential generated by the various
factors cancels, leading to a branch with $W=0$~\cite{GA-PRL}.  In theories
where anomalies do not match at the origin, $G_P$ is such that it is impossible
to have a $W=0$ branch, and the theory always has a non-zero dynamical
superpotential. When $\mu = \mu_{adj}$, all free algebra theories have unbroken
$U(1)$ gauge symmetries in the bulk of the moduli space, a fact recently
noticed in Ref.~\cite{CW}. In this case, after including photons and sometimes
also massless monopoles in the low-energy spectrum, anomalies are saturated at
the origin. For $\mu > \mu_{adj}$, we argue based on flows that the free
algebra theories are in an interacting non-Abelian Coulomb phase at the
origin of the moduli spaces. Theories with $\mu$ $>$, $<$, or $=$ $\mu_{adj}$
can only flow via the Higgs mechanism to other theories of the same type.

Free algebra theories have been classified in the mathematics
literature~\cite{Schwarz,AG}. After selecting theories free of gauge anomalies
it turns out that all $\mu < \mu_{adj}$ examples have no constraints among the
basic invariants~\cite{GA-PRL}. The requirement that the theory have no gauge
anomalies is essential for the result, which is why this fact was not noticed
earlier in the mathematics literature. We also recall little
known theorems proven in
the mathematics literature which are helpful in analyzing supersymmetric gauge
theories. First, theories that have unbroken $U(1)$'s in the bulk of the moduli
space must have $\mu = \mu_{adj}$~\cite{Elashvili}. Second, any theory with
$\mu > \mu_{adj}$ has completely broken gauge group in the bulk of the moduli
space~\cite{AnViEh}.

In the next section, we analyze the low-energy dynamics of the free algebra
theories. A particularly interesting example studied in detail is the $SO(13)$
gauge theory with a spinor. This is a theory with $W\not=0$ which flows via
the Higgs mechanism to a theory with a $W=0$ branch. A complete list of free
algebra theories with $\mu < \mu_{adj}$ is presented in Appendix A.
There, we also indicate patterns of
gauge symmetry breaking for these theories. In Appendix B, we discuss more
mathematical issues. We derive a formula for the dimension of the classical
moduli space, which is found to be equal to the number of microscopic degrees
of freedom minus the dimension of the gauge group plus the dimension of the
smallest unbroken subgroup. In applying this formula, one needs to be careful
about the possibility that the smallest unbroken subgroup at a D-flat point is
not the same as the smallest unbroken subgroup preserved by an arbitrary field
configuration. This  possibility occurs for $SO(10)$ gauge group with a single
spinor field and the $SU(2 N+1)$ theory with an antisymmetric tensor and $2
N-3$ antifundamentals.

\section{Free Algebra Theories\label{sec:FA}}
\setcounter{equation}{0}
The moduli space of supersymmetric gauge theories can be parameterized by the
vacuum expectation values (VEVs) of gauge invariant polynomials made out of
matter superfields. For any theory there is a minimal set of invariants. In
general, the basic gauge invariants are not independent, but are subject to
nonlinear constraints. The VEVs allowed by these constraints are in one-to-one
correspondence with the D-flat configurations of the microscopic degrees of
freedom. Some theories, however, have unconstrained basic gauge invariants. 
These theories are the subject of this paper. We give a precise definition of
classical moduli space and free algebra of invariants in Appendix~B, where
we also explain how to calculate the dimension of the moduli space.

All theories with a free algebra of invariants together with their basic set of
invariants are given in Refs.~\cite{Schwarz,AG}. We restrict our attention to
physical theories, that is those which are free of gauge anomalies. Since the
dynamics of the free algebra theories with anomaly matching between the
microscopic fields and the basic gauge invariants was already
studied~\cite{GA-PRL}, we now consider only cases where anomalies do not match.
We present  all the cases in the following sections, dividing them according to
the kind of non-perturbative effects that take place.

We will frequently consider theories along a flat direction, which partially
or completely breaks the gauge group. It is easy to see that the sign of 
$\mu-\mu_{adj}$ is preserved under such Higgs flow. In fact,
when $G$ is broken to a subgroup $H$, $n (\mu - \mu_{adj}) = \tilde{\mu} -
\tilde{\mu}_{adj}$, where $n$ is an integer, $\mu - \mu_{adj}$ is computed
in $G$, and $\tilde{\mu} - \tilde{\mu}_{adj}$ is computed in $H$.
The value of $n$ depends on the embedding of $H$ in $G$, and it is the $G$
winding number of the unit $H$ instanton.

\subsection{$\mu < \mu_{adj}$}

It is a straightforward, yet tedious, exercise to select all physical theories
with $\mu < \mu_{adj}$ among the free-algebra models of Ref.~\cite{Schwarz,AG}.
We list all of these theories in Appendix~A. For each theory we give a sample
pattern of symmetry breaking and the smallest unbroken subgroup along a generic
flat direction. It turns out that there are no other $\mu < \mu_{adj}$ theories
besides the ones with a free algebra~\cite{GA-PRL}.  Such statement is not true
for theories with gauge anomalies. For example, an unphysical $SU(N)$ theory
with $(N+2)$ fields in the fundamental representation and no fields in the
antifundamental representation has constraints among the basic gauge
invariants. The basic gauge invariants are $B_{i_1 \ldots i_N}= \det Q_{i_1}
\ldots Q_{i_N}$, which obey the constraint equations $B_{i_1 \ldots i_N}
B_{i_{N+1} j_1 \ldots j_{N-1}} \epsilon^{i_1 \ldots i_{N+2}}=0$.

$U(1)$ global symmetries, including the R-symmetry, restrict the form of 
a dynamically generated superpotential. In terms of the
microscopic matter fields, $\phi_i$, the superpotential can be written
as~\cite{ADS}
\begin{equation}
  W_{dyn} \propto \left( \frac{\Lambda^{(3 \mu_{adj}-\mu)/2}}{\prod_i
  \phi_i^{\mu_i}} \right)^{\frac{2}{\mu_{adj}-\mu}},
\end{equation}
where $\mu_i$ is the Dynkin index of the $i$-th representation and $\mu=\sum_i
\mu_i$. Whenever $\mu < \mu_{adj}$ such a superpotential has a good classical
behavior. $\Lambda$ occurs in the numerator, so that $W_{dyn} \to 0$ for large
values of the fields.
For $\mu=\mu_{adj}$, R symmetry prohibits the generation
of a superpotential. For $\mu > \mu_{adj}$, a dynamical superpotential
cannot be generated in the free algebra theories, since $\Lambda$ occurs in the
denominator, and $W_{dyn}$ does not have the correct
behavior for large values of the fields. In those cases where
a quantum superpotential is allowed, one still needs to check whether or
not it is actually generated by non-perturbative effects. One way of checking
is by integrating out matter fields from theories with a larger number of
flavors using the results of Refs.~\cite{s-conf,BD,Peter,CW}. This approach
can frequently be very cumbersome as theories with additional
matter fields tend to have a large number of basic gauge invariants and
complicated superpotentials. Here, we take a different path. We explore points
on the moduli space where the original gauge group is broken. The effective
theory below the scale of the massive gauge bosons is frequently known, and
is usually easier to analyze than the original theory. 

All $\mu<\mu_{adj}$ theories are listed in Appendix A. For those analyzed
in Ref.~\cite{GA-PRL} for which anomalies match at the origin, the smallest
unbroken subgroup at generic points on the moduli space turns out to be
a product group. The remaining pure Yang-Mills theory
exhibits gaugino condensation and has superpotentials generated in each
factor group. The branch with zero superpotential arises due to a cancellation
between the contributions from different factor groups to the superpotential.

We now explain why anomalies do not match for the remaining theories. They are
listed in Appendix A, and can be divided into three types:
({\it i}) the gauge group is
completely broken, ({\it ii}) it is broken to a simple group, ({\it iii})
it is broken to a semi-simple group. For example, $SU(N)$ with $N-1$ flavors
is an example of case ({\it i}); if the number of flavors is
smaller than $N-1$, it is an example of $({\it ii})$. 
$SU(2 N+1)$ theory with an antisymmetric tensor and its conjugate is 
an example of ({\it iii}), since the gauge group is broken
down to $SU(2)^N$ at generic points on the moduli space.

Inspecting the table in Appendix A we conclude that in case ({\it i})
the Higgs flow always contains an $SU(2)$ group with two fields in
the fundamental representation. The $SU(2)$ theory with two fundamentals
has been shown by an explicit calculation to have a non-zero superpotential
generated by instantons~\cite{ADS}. Using scale-matching relations we
checked that this has to be the case for all other theories with completely
broken gauge group at generic values of the moduli fields.

In case ({\it ii}), when the maximal breaking leaves a simple group, we can
also use the scale
matching to show that there is a non-zero superpotential. From the study
of supersymmetric QCD~\cite{Seiberg} and other simple groups with fundamental
matter fields~\cite{SO,Sp,G2} we know that pure ${\cal N}=1$ Yang-Mills
theories exhibit gaugino condensation, which is responsible for the generation
of a superpotential. A large fraction of theories listed in Appendix A
have a non-zero superpotential due to gaugino condensation in the unbroken
subgroup. We comment on some examples from this class at the end of this
section.

The remaining case is when semi-simple groups remain unbroken at the
points of maximal breaking. Obviously,
gaugino condensation takes place in each unbroken factor, but there is
no guarantee that contributions to the superpotential from the
different factor groups do not cancel. After all, this effect was
responsible for vanishing superpotentials in theories with anomaly matching.
There are only seven theories without anomaly matching which have unbroken
product groups at generic points on the moduli space.
These are $SO(14)$ with $2\, \Yfund + S$, $SO(14)$ with $\Yfund + S$,
$SO(13)$ with  $\Yfund + S$, $SO(13)$ with $S$, $SO(12)$ with $S+S'$,
$SU(6)$ with $\Ythreea +\Yfund + \overline{\Yfund}$ and
$SU(2 N+1)$ with $\Yasymm+\overline{\Yasymm}$.\footnote{$S$ and $S'$ 
denote the spinor representations.} (In the table in Appendix A,
these theories are numbered 53, 53, 30, 31, 52,
11 and 6, respectively.) The behavior of the different theories is related,
since there are flows between them along certain flat directions:
\begin{eqnarray}
SO(14): 2\, \Yfund + S &
   \stackrel{\langle\, \Yfund\, \rangle}{\longrightarrow} &
   SO(13): \Yfund + S \stackrel{\langle\, \Yfund\, \rangle}{\longrightarrow}  
   SO(12): S+S' \stackrel{\langle S \rangle}{\longrightarrow}
   SU(6): \Ythreea + \Yfund + \overline{\Yfund}
   \stackrel{\langle\, \Yfund\, +\, \overline{\Yfund}
             \, \rangle}{\longrightarrow}  
   \nonumber \\
&\longrightarrow& SU(5): \Yasymm+\overline{\Yasymm}
      \stackrel{\langle\, \Yasymm\, +\, \overline{\Yasymm}
                \, \rangle}{\longrightarrow}
      SU(2)\times SU(2),
\nonumber \\
SO(14): \Yfund + S & \stackrel{\langle\, \Yfund\, \rangle}{\longrightarrow} &
   SO(13): S  \stackrel{\langle S \rangle}{\longrightarrow} SU(3)\times SU(3),
\nonumber
\end{eqnarray}
where we indicated the smallest unbroken subgroup as the last step of the
flow. We only need to understand why non-zero superpotentials are generated in
the $SU(2N+1)$ theory with $\Yasymm+\overline{\Yasymm}$ and the $SO(13)$
with a spinor; all other cases are explained by scale matching along flat
directions.

It is interesting that $SU(2N)$ with $\Yasymm+\overline{\Yasymm}$ has
a branch with a zero superpotential due to cancellations among gaugino
condensates in the unbroken $SU(2)^N$ subgroup~\cite{GA-PRL}. On the other
hand, $SU(2N+1)$ does not have a vanishing superpotential, even though
it has the same $SU(2)^N$ unbroken subgroup. The VEV of the tensor fields
which breaks $SU(2N+1)$ to $SU(2)^N$ is 
\begin{equation}
  \langle\, \Yasymm \,\rangle =\langle\, \overline{\Yasymm}\, \rangle =
      {\rm diag}(v_1 \sigma_2,v_2 \sigma_2,\ldots,v_N \sigma_2,0).
\end{equation}
The massive gauge bosons, which enter the scale matching, decompose
under $SU(2)^N$ as the $2 \cdot (2,2,1,\ldots,1) (+\ {\rm permutations})$
and $2 \cdot (2,1,1,\ldots,1) (+\ {\rm permutations})$. 
The masses squared of the $(2,2,\ldots)$ bosons are proportional to
$(v_i-v_j)^2$ for one set and to $(v_i+v_j)^2$ for the other. In case of the
the $(2,1,\ldots)$ bosons the masses squared  are proportional to $v_i^2$.
Thus, the scale of the $i$-th $SU(2)$ factor is related to the scale $\Lambda$
of $SU(2N+1)$ by
\begin{equation}
  \Lambda^6_i v_i^2 \prod_{j \neq i} (v_i^2-v_j^2)^2 = \Lambda^{4N+4},
\end{equation}
and the dynamical superpotential is
\[
W = \sum_i \pm \Lambda_i^3.
\]
For simplicity, let us examine the case of the $SU(7)$ group. The
superpotential induced by gaugino condensation is
\begin{equation}
  W=\Lambda_{SU(7)}^8 \left[ \pm \frac{1}{v_1 (v_1^2-v_2^2) (v_1^2-v_3^2)}
                             \pm \frac{1}{v_2 (v_2^2-v_1^2) (v_2^2-v_3^2)}
                             \pm \frac{1}{v_3 (v_3^2-v_1^2) (v_3^2-v_2^2)}
                      \right],
\end{equation}
which does not have a zero branch.  For $SU(2N)\rightarrow SU(2)^N$,
there are no massive gauge
bosons in the $(2,1,\ldots)$ representation, and matching involves
only products of $(v_i^2-v_j^2)$. This combination leads to a cancellation,
as pointed out in Ref.~\cite{GA-PRL}.

Now let us turn to the $SO(13)$ theory with one field $S$ in the spinor
($\bf{64}$) representation. A particular VEV of the spinor field can break
$SO(13)$ to an $SU(6)$ theory with one field in the three-index
antisymmetric representation. This $SU(6)$ theory with $T=\Ythreea$ has been
studied in Refs.~\cite{s-conf,GA-PRL} and it has two branches. One branch
has $W=0$, another $W={\Lambda^5_{SU(6)}}/{\sqrt{T^4}}$.
Since anomalies do not match in the $SO(13)$ theory, it
must have a non-vanishing superpotential that somehow gives both $W=0$ and
$W={\Lambda^5_{SU(6)}}/{\sqrt{T^4}}$ branches along the flat direction
which breaks $SO(13)$ to $SU(6)$.

Fortunately, patterns of symmetry breaking in $SO(13)$ group with a spinor
field have been classified in Ref.~\cite{so13}. There are two gauge
invariants in this theory: $X=S^4$ and $Y=S^8$, so there are two inequivalent
flat directions. Let us parameterize these flat directions at the microscopic
level by two inequivalent spinor expectation values $S_1$ and $S_2$.
We describe the moduli space as a linear combination $\alpha S_1 + \beta S_2$.
In terms of these parameters
$X=\alpha^4 + \beta^4$ and $Y=\alpha^4 \beta^4$~\cite{so13}.
Depending on the relative values of $\alpha$ and 
$\beta$ there are different patterns of $SO(13)$ breaking. For generic values
of $\alpha$, $\beta$ $SO(13)$ breaks to $SU(3)\times SU(3)$ with no matter
fields transforming under the unbroken gauge group. When $\alpha=0$ and
$\beta \neq 0$, $SO(13)$ is broken to $SU(6)$ with $\Ythreea$. Finally,
when $\alpha=\beta \neq 0$, $SO(13)$ is broken to $SU(3) \times G_2$
with one field transforming as $(1,7)$ under $SU(3) \times G_2$.

{}From the previous analysis~\cite{GA-PRL} of $SU(6)$ with $\Ythreea$ we know
the superpotential in the $SU(3)\times SU(3)$ theory along a generic flat
direction. The $SU(3)\times SU(3)$ is a product of two independent Yang-Mills
theories in which gaugino condensation takes place. However, the
superpotential has a relative minus sign between the two contributions
from the $SU(3)$ factors~\cite{GA-PRL}:
$W=\omega^i \Lambda_1^3 - \omega^j \Lambda_2^3$,
where $\omega$ is the cube root of unity, $i,j=1,2,3$ and
$\Lambda_{1,2}$ are the characteristic scales of each $SU(3)$ factor.
Therefore, using scale matching in $SU(3)\times G_2$ with $(1,7)$ we get
\begin{equation}
  W=\omega^i \Lambda_1^3 -
          \omega^j \left(\frac{\Lambda^{11}_2}{Q^2}\right)^{1/3}
\end{equation}
along the $\alpha=\beta\neq 0$ flat direction. In the above equation
$Q$ denotes the $\bf 7$ of $G_2$ and $\Lambda_2$ the scale of $G_2$.
Knowing that $SO(13)$ is broken to $SU(3)\times G_2$ when $\alpha=\beta$, we
can identify the field $Q$ in the $(1,7)$ representation of $SU(3)\times G_2$
with $\alpha-\beta$. We can also use the breaking of $SO(13)$ to $SU(6)$
and then to $SU(3)\times SU(3)$ to match $\Lambda_1$ with the $SO(13)$ scale:
$\Lambda^{25}=\Lambda_1^9 X Y^{3/2}$. Combining all of this information
together we obtain the superpotential for the full $SO(13)$ theory
\begin{equation}
  W=\frac{\Lambda^{25/3}}{X^{1/3} Y^{1/2}} \left(\omega^i -
    \omega^j \frac{X^{2/3}}{(X^2-4 Y)^{1/3}} \right).
\end{equation}
In the above superpotentials we explicitly display the cube roots of unity
rather than  hide them  in the definition of the cube root.
When $SO(13)$ is broken to $SU(6)$ at a scale much larger than that of
$SU(6) \to SU(3) \times SU(3)$, then $\alpha \ll \beta$, so that 
$Y \ll X^2$. In this limit $W \approx
({\Lambda^{25/3}}/{X^{1/3} Y^{1/2}}) (\omega^i -\omega^j)$.
Indeed this superpotential correctly reproduces the superpotential
of the $SU(6)$ theory. When $i=j$ we obtain the zero branch, and for
$i \neq j$ the branch where $W \propto {\Lambda_6^5}/{\sqrt{T^4}}$.

In the remainder of this section we consider again theories with a simple
unbroken gauge group at generic points of the moduli space. Several cases
require a more careful analysis due to the presence of instantons in the
broken group. For a detailed discussion of this effect see
Ref.~\cite{CH}.~\footnote{We thank C. Cs\'aki for discussions about this
topic.} When a one-instanton configuration in the unbroken subgroup
does not correspond to one instanton in the original group there can be
additional contributions to the superpotential from the instantons not
represented in the unbroken subgroup. Among the theories with
$\mu < \mu_{adj}$ this can happen in $SO(N)$ with $N-3$ vectors~\cite{SO},
$SU(7)$ with $\Ythreea + 2\, \overline{\Yfund}$, and $Sp(6)$ with
$\Ythreea + \Yfund$. The particular flat directions for the $SU(7)$ and
$Sp(6)$ theories listed in Table 3 of Appendix A yield subgroups with
the index of embedding equal to two. Both theories have an unbroken $SU(2)$
at generic points of their moduli spaces, thus one expects
superpotentials generated by gaugino condensation. However, the instanton
contribution from the broken part of the gauge group could potentially
cancel the contribution arising due to gaugino condensation.

This is indeed what happens in the case of $SO(N)$ with $N-3$
vectors~\cite{SO}. The $SO(N)$ theory has two branches. One branch has
a dynamically generated superpotential, and another one has additional massless
states at the origin of the moduli space. Those new massless particles are
needed to saturate anomalies. We will argue that this does not happen
for the $SU(7)$ and $Sp(6)$ theories, which only have
branches with non-zero dynamical superpotentials. One way of obtaining the
superpotentials for these theories would be a direct computation of the
coefficient of the instanton contribution to the superpotential. Instead,
we explore different regions of the moduli spaces of these theories, where
there are no non-trivial instanton configurations in the broken part of
the group.

Apart from the breakings that we list in Table 3, there are other special
points of enhanced symmetry on the moduli spaces of the $SU(7)$ and $Sp(6)$
theories. The invariants for these theories were given in Ref.~\cite{BD}.
They are $B^7$ and $B^3 \bar{Q}^2$ for the $SU(7)$ theory, $B^4$ and
$B^2 Q^2$ for $Sp(6)$, where $B$ denotes the three-index antisymmetric tensor
and $Q (\bar{Q})$ denote fields in the (a-)fundamental representation.
The symmetry breaking with a non-trivial index of embedding arises when 
the fields $B$ obtain a VEV with the other fields having zero VEVs, which
corresponds to nonzero values of $B^7$, $B^4$, respectively. The $SU(7)$
symmetry is also enhanced at points where $\langle B^7 \rangle=0$,
$\langle B^3 \bar{Q}^2 \rangle \neq 0$. Similarly $\langle B^4 \rangle=0$,
$\langle B^2 \bar{Q}^2 \rangle \neq 0$ is an enhanced symmetry point
for $Sp(6)$. Such choices of VEVs give the following unbroken subgroups
\begin{eqnarray}
  SU(7): \Ythreea + 2\, \Yfund & \rightarrow & Sp(6): \Ythreea + \Yfund,
     \nonumber \\
  Sp(6): \Ythreea + \Yfund & \rightarrow & SU(2)\times SU(2): (\Yfund,\Yfund).
     \nonumber
\end{eqnarray}
These breakings do not have extra instanton contributions since the index of
embedding is one. Moreover, since the $SU(7)$ theory flows to $Sp(6)$,
we need only to analyze the latter one.

We now argue that $Sp(6)$ with $\Ythreea + \Yfund$ has only branches with
non-zero dynamically-generated superpotentials. Consequently, the same result
applies to $SU(7)$. As we mentioned earlier, along the $B^2 Q^2$ flat
direction $Sp(6)$ is broken to $SU(2)\times SU(2)$ with one field
in the $(\Yfund,\Yfund)$ representation. Such a theory has been described in
Ref.~\cite{su2su2}, and was found to have the superpotential
$W=(\Lambda_1^{5/2}\pm \Lambda_2^{5/2})^2 / P^2$, where $P$ denotes the
field in the $(\Yfund,\Yfund)$ representation, and $\Lambda_i$'s are the
scales of the $SU(2)$ groups.  At the microscopic level, the
VEV which breaks $Sp(6)$ to $SU(2)\times SU(2)$ can be chosen to be
$B_{245}=v$, $B_{346} =-v$ and $Q_1=v_Q$ with all other independent components
of $B$ and $Q$ set to zero. $B_{245}$ and $B_{346}$ need to have opposite
signs due to the tracelessness requirement for irreducible representations
of $Sp$. The unbroken $SU(2)$'s act on the $(2,5)$ and $(4,6)$ indices of the
original $Sp(6)$. Of course, the D-flatness condition relates $v$ and $v_Q$,
but we prefer to distinguish them for the time being. 

Scales $\Lambda^5_{1,2}$ of the two $SU(2)$ factors have the same magnitude.
They are, however, opposite in sign due to a discrete interchange symmetry
of the two $SU(2)$'s. A similar relative sign of two scales was found in an
$SU(6)$ theory broken to $SU(3)\times SU(3)$~\cite{GA-PRL}. In terms of the
VEVs and the $(\Yfund,\Yfund)$ field $B^4=v^2 P^2$, $B^2 Q^2=v^2 v_Q^2$ and 
\begin{equation}
  \frac{\Lambda^9_{Sp}}{B^4 \sqrt{B^2 Q^2}} =
  \frac{\Lambda^9_{Sp}}{P^2 v^3 v_Q}=\frac{\Lambda_{1,2}^5}{P^2},
\end{equation}
where the first term is the most general $Sp(6)$ superpotential consistent
with the symmetries. Under the interchange of the two $SU(2)$'s we have
$v_Q \rightarrow v_Q$,
but $v \rightarrow -v$. Therefore, the two $\Lambda_i^5$'s must differ by
a minus sign. This means that the $SU(2)$ theories have a relative $\theta$
angle equal to $\pi$. Consequently, the superpotential generated in the
unbroken $SU(2)\times SU(2)$ is
$W=\Lambda_{Sp}^{9}(1 \pm i)^2 / (B^4 \sqrt{B^2 Q^2})$, which does not
vanish for either choice of the sign in the numerator.
Thus, we have shown that the $Sp(6)$ and $SU(7)$ theories have only
branches with non-zero superpotentials. Since the form of the superpotential
is unique in these theories, one can use the superpotential obtained this way
and examine flat directions directions with potential instanton effects,
thus obtain the coefficient of the instanton contribution in the broken part
of the group.

\subsection{$\mu = \mu_{adj}$}
\label{sec:mueq}
The list of all free algebra theories with $\mu = \mu_{adj}$ is quite short.
It includes all theories with an adjoint superfield, which are automatically
${\cal N}=2$ Yang-Mills theories, and also
$SO(N)$ with $N-2$ vectors, $SU(6)$ with 2 $\Ythreea$ and $Sp(6)$ with 2
$\Yasymm$. One can check that anomalies are not saturated by the basic gauge
invariants in any of these theories.

One can also check that at generic points of the moduli space there are
only unbroken $U(1)$ gauge symmetries. All these theories are therefore
in the Abelian Coulomb phase at low energies. Since the low-energy spectrum
includes $U(1)$ photons their contribution has to be included in anomaly
matching. Anomalies at the origin indeed match after including photons,
the only exception being the $SO(N)$ theory with $N-2$ vectors, which has
massless monopoles at the origin~\cite{SO}.
For all these theories the low-energy dynamics has been determined, namely the
Seiberg-Witten curves have been found. For a list of references see~\cite{CW}.

It is interesting that the presence of unbroken $U(1)$'s is specific to the
$\mu=\mu_{adj}$ case. A theorem by \'Elashvili~\cite{Elashvili} states
that for a simple
group if there are unbroken $U(1)$ symmetries in the bulk of the moduli space
then $\mu=\mu_{adj}$. The bulk of the moduli space means here that the
set of points where the gauge group is broken to a product of $U(1)$'s
is open (and dense) in the  Zariski topology. We recall the definition
of Zariski topology in Appendix~B. The fact about
$U(1)$'s in the bulk of the moduli space was also noticed in Ref.~\cite{CW}.

As we already mentioned none of the free algebra theories with $\mu=\mu_{adj}$
has anomalies saturated by the basic gauge invariants. It is interesting
that the same is true for the $\mu=\mu_{adj}$ theories with constraints. 
For the $\mu=\mu_{adj}$ constrained theories, the low-energy spectrum does
consist of the basic gauge invariants, but some of the constraints are quantum
modified. Due to the modification, some fields can be eliminated from the
theory, and it may also happen that the modification excludes the
origin from the quantum moduli space~\cite{Seiberg,BD,Peter}.
Note that for s-confining theories~\cite{s-conf} integrating out fields such
that the effective theory has $\mu=\mu_{adj}$ always gives theories with
a quantum modified moduli space.

\subsection{$\mu > \mu_{adj}$}
First, we want to point out a general result about  $\mu > \mu_{adj}$
theories due to Andreev, Vinberg and \'Elashvili~\cite{AnViEh}.
At a generic point of the moduli space the gauge group is completely broken,
that is at most discrete symmetries remain unbroken. This theorem extends to
all theories with semi-simple groups with $\mu > \mu_{adj}$, not only the
ones with a free algebra of invariants. For a gauge group which is a product
of simple groups, the group is completely broken in the bulk of the
moduli space if each factor satisfies $\mu > \mu_{adj}$.

The number of free-algebra theories with $\mu > \mu_{adj}$ is small.
Those with anomaly matching were described in Ref.~\cite{GA-PRL}.
There are three remaining ones where anomalies do not match:
$SU(N)$ with $\Ysymm + \overline{\Ysymm}$, $Sp(6)$ with 2 $\Ythreea$ and
$SO(N)$ with $\Yasymm + \Yfund$. We list the gauge invariants of these
theories in Table~\ref{table:inv} for completeness.

\begin{table}[tb]
\renewcommand{\arraystretch}{1.5}
\begin{center}
\begin{tabular}{|l|l|l|} \hline
  $SU(N)$ & $\Ysymm + \overline{\Ysymm} $ &
      $Tr(S \overline{S})^i$, $i=1,\ldots,N-1$ \\ 
  $Sp(6)$ & $2\, \Ythreea$ & $A_1 A_2$; $A_1^3 A_2^3$;
                                  $A_1^i A_2^{4-i}$, $i=0,\ldots,4$  \\  
  $SO(2 N)$ & $\Yasymm+\Yfund$ & $A^{2 i}$, $i=1,\ldots,N-1$;
                                       $A^{2 k} Q^2$, $k=0,\ldots,N-1$;
                                       $A^N$ \\ 
  $SO(2N+1)$ & $\Yasymm+\Yfund$ & $A^{2 i}$, $i=1,\ldots,N$;
                                        $A^{2 k} Q^2$, $k=0,\ldots,N-1$;
                                        $A^N Q$ \\ \hline
\end{tabular}
\end{center}
\caption{\label{table:inv}
Invariants of theories with $\mu > \mu_{adj}$ for which anomalies
do not match at the origin. We indicate the gauge group in the first column,
the field content in the second  and the invariants in the last column.
In writing the invariants, the fundamental representation is denoted by $Q$,
symmetric tensors by $S$ and antisymmetric tensors by $A$. A subscript
distinguishes between the two different $\protect\Ythreea$.}
\end{table}

As usual, we explore the moduli space of these theories. $Sp(6)$ with 2
$\Ythreea$ flows to $SU(3)$ with $\Ysymm + \overline{\Ysymm}$ after giving
a VEV to the tensor fields. This is a particular case of $SU(N)$
with $\Ysymm + \overline{\Ysymm}$, which flows to an $SO(N)$ with $\Ysymm$
when one of the tensors gets a VEV. The $SO(N)$ theory with $\Ysymm$ has
been recently discussed in Ref.~\cite{BCI}. The authors have argued that
at the origin of the moduli space this theory is in a non-Abelian Coulomb
phase. The same must then also be true for the
$SU(N)$ with $\Ysymm + \overline{\Ysymm}$ and $Sp(6)$ with 2 $\Ythreea$.

The remaining theory $SO(N)$ with $\Yasymm + \Yfund$ is an example of a theory
with a chiral superfield in the adjoint representation and additional matter
fields, but no tree level-superpotential. The description of these kinds of
theories is only known after the moduli space is restricted by adding a
tree-level superpotential~\cite{KSS,ILS}.

\section{Conclusions\label{sec:Conclusions}}
We have studied all ${\cal N}=1$ theories without constraints among the
basic gauge invariants. These include all theories with $\mu < \mu_{adj}$,
theories in the Coulomb phase for $\mu = \mu_{adj}$~\cite{CW}, and a few
examples with $\mu > \mu_{adj}$. There are only two kinds of low-energy
dynamics in all $\mu < \mu_{adj}$ theories. When anomalies match at the
origin, the theory has a confining branch with no dynamical superpotential
and branches with a dynamically-generated superpotential~\cite{GA-PRL}.
When anomalies do not match there is always a dynamically-generated
superpotential, which excludes the origin from the quantum moduli space.
We have checked that for each theory a superpotential is indeed generated.
This task was accomplished by studying the flows of these theories
along flat directions. We give a complete list of all $\mu < \mu_{adj}$
theories together with a pattern of symmetry breaking that was helpful
in analyzing each case in Appendix A. It is interesting that for most simple
groups, the smallest unbroken subgroup is not restricted by the D-flatness
condition.
With the exception of $SO(10)$ with a spinor field and $SU(2N+1)$ with
$2 N-3$ antifundamentals, a generic field configuration preserves the
same gauge symmetries as the most general D-flat configuration
(see Appendix B).

Some results obtained by mathematicians aid in the understanding of the
$\mu \geq \mu_{adj}$ cases. Theories with the $U(1)$ gauge bosons in the
bulk of the moduli space can occur only when $\mu = \mu_{adj}$. For
$\mu > \mu_{adj}$ generic flat directions turn out to completely break
the gauge group.

Table~\ref{tab:summary} summarizes all possibilities for ${\cal N}=1$
theories. The low energy dynamics of all theories with $\mu \leq \mu_{adj}$
is now understood. The last column in the table is mostly
uncharted territory. There are only a few examples of theories whose dynamics
is known to be described in terms of dual theories. Several examples
with $\mu = \mu_{adj}+2$ are known to exhibit s-confinement.
For all other $\mu > \mu_{adj}$ theories there is as yet no
systematic analysis, and it is possible that new non-perturbative phenomena
remain to be discovered.

\begin{table}[tb]
\begin{tabular}{|r@{}l|c|c|c|} \hline
 & & $\mu < \mu_{adj}$ & $\mu = \mu_{adj}$ & $\mu > \mu_{adj}$
       \\ \hline \hline 
 & irred & $SO(14)$ $S$ & --- & $SO(N)$ $\Ysymm$ \\ 
\raisebox{1.5ex}[0pt]{FA $\surd$ $\Big\{$} & red & $SU(N)$
      $\Yasymm + \overline{\Yasymm}$ & --- & --- \\
 & irred & $SO(N)$ $\Yfund$ & $SU(N)$ $adj$ & --- \\
\raisebox{1.5ex}[0pt]{FA $X$ $\Big\{$} & red &
      $SU(N)$ $\Yfund+\overline{\Yfund}$ &
      $SO(N)$ $(N-2)\, \Yfund$ & $SO(N)$ $(N-1)\, \Yfund$ \\ 
 & irred & --- & --- & ---  \\
\raisebox{1.5ex}[0pt]{CN $\surd$ $\Big\{$} & red & --- & --- & 
         $SU(N)$ $(N+1) \, (\Yfund + \overline{\Yfund})$ \\
 & irred & --- & --- & $SO(15)$ $S$ \\
\raisebox{1.5ex}[0pt]{CN $X$ $\Big\{$} & red & --- &  
         $SU(N)$ $ N \, (\Yfund + \overline{\Yfund})$ & 
         $SU(N)$ $ (N+2) \, (\Yfund + \overline{\Yfund})$\\ 
\hline
\end{tabular}
\caption{  \label{tab:summary} Summary of all possible
${\cal N}=1$ theories with simple groups. The rows of the table
divide theories according to the algebra of invariants (FA=free algebra,
CN=constraints), anomaly matching ($\surd$) or lack of it ($X$)
and reducibility of the gauge representation of the microscopic field.
We indicate by a dash that there are no examples of theories of a given kind.
When there exist theories in a given class we give an example. $S$ indicates
the spinor representation for $SO$ theories.}
\end{table}

\section*{Acknowledgements}
We thank  Ken Intriligator and Erich Poppitz for discussions.
This work was supported in part by
the U.S. Department of Energy under contract DE-FG03-97ER405046.

\section*{Appendix A \hspace*{0.25cm} Flows of $\mu < \mu_{adj}$ theories}
\setcounter{equation}{0}
\renewcommand{\theequation}{A.\arabic{equation}}
All supersymmetric gauge theories with simple gauge 
group, zero tree-level superpotential and $\mu < \mu_{adj}$ 
have a free algebra of gauge invariant operators. 
In the table below we present all these theories, together with their 
Higgs flow along selected flat directions and the smallest unbroken subgroup.
Theories are listed in the second column, and indicated by their
gauge group followed by matter content and comments.
For convenience, the theories are numbered in the first column,
except for those theories with anomaly matching at the origin, 
for which the entry number has been replaced with {\bf A}, and 
those with a non-reductive stabilizer (Appendix B), 
by ${\bf NR}$. The fourth column gives the unbroken subgroup
and its field content along the flat direction associated with the
field listed in the third column. We omit gauge singlets in the field
content of the theories listed in the fourth column.
{\bf TDF} in the third column
means that the only solution to the D-flatness condition is the
trivial one. Since by breaking a $\mu < \mu_{adj}$ theory one arrives
at another $\mu < \mu_{adj}$ theory, one can follow the pattern
of symmetry breaking using the table. The entry number
of the unbroken subgroup is listed in the fifth column. The final result
of the the flow---the smallest unbroken subgroup---appears in the last
column of the table.

Our notation is the following. $\phi_k$ denotes a $k$-index totally
antisymmetric tensor (for $Sp$ groups its highest weight irreducible
component).  For example, $\phi_1$ means the fundamental representation,
$\phi_2$ means $\Yasymm$, etc. $S$, $S'$ denote the spinor representations,
and an asterix indicates the conjugate representation. The variable
$s_i$ takes integer values $1,2,\ldots,i$. When relevant, we indicate even
numbers with subscript $e$, and the odd ones with $o$. If the smallest
unbroken subgroup is trivial we indicate it by $[\;\;]$. It should also
be understood that $SU(4-s_3)$ means $[\;\;]$ when $s_3=3$, etc.
The group isomorphisms $SO(6) \cong SU(4)$, $SO(5) \cong Sp(4)$,
$SO(3) \cong SU(2)$ and the outer automorphisms of $SO(8)$ which
permutes $\p1$, $S$ and $S'$ were used to avoid redundant entries.
In a few cases where the gauge group of a theory was 
omitted due to space limitations, it is understood to be
the same as the group in the table entry right above it.

\begin{small}
\setlongtables
\setlength{\tabcolsep}{1mm}
\renewcommand{\arraystretch}{1.3}
\newcolumntype{L}{>{$}l<{$}}
\newcolumntype{C}{>{$}c<{$}}
\begin{longtable}{|L|L|C|L|C|L|} 
\caption{All supersymmetric gauge theories with $\mu < \mu_{adj}$}\\
\hline 
\endhead 
\hline 
\multicolumn{6}{r}{{ \slshape
 continued on the next page}} \\ 
\endfoot
\hline 
\multicolumn{6}{r}{{ \slshape 
   That's all folks}} \\
\endlastfoot
\hline
&\mathbf{SU(n)}\;\; ${\bf theories}$  &<\;>&${\bf Higgs flow}$&
$entry$& ${\bf SYM}$\\ \hline 
(1) & k(\p1 + \ps1):k<n & \p1+\ps1 & SU(n-1):(k-1)(\p1+\ps1)&
    (1)  &  SU(n-k)\\
(2) & \p2+s_2\p1+(n-4+s_2)\ps1 & \p1+\ps1& \p2+s_2\p1+ (n-5+s_2)\ps1 &
      (1,2)&SU(3-s_2)\\
(3) & \p2+(n-4)\ps1:n_e & 
     2\ps1+\p2 & SU(n-2):\p2+(n-6)\ps1 &(3,32)& SO(5)\\
{\bf  NR}&\p2+(n-4)\ps1:n_o &  
    2\ps1+\p2 &  SU(n-2): \p2+(n-6)\ps1 & (4,10) &SU(5) \\
(5) &\p2 + \ps2 +\p1+\ps1 & \p1+\ps1 &  SU(n-1):\p2+\ps2+\p1 +
     \ps1 & (1,5) & [\;\;] \\
(6) &\p2+\ps2:  n_o \geq 5&
    \p2+\ps2 &&& \left(SU(2) \right)^{(n-1)/2} \\
{\bf A}&\p2 + \ps2: n_e\geq 4&
    \p2+\ps2 &&& \left(SU(2) \right)^{n/2} \\
(8) & SU(4): 2\p2+\p1+\ps1 & \p1+\ps1 &  SU(3): 2\p1+2\ps1 &(1) & [\;\;] \\
(9) &  SU(5): 2(\p2+\ps1) &  2\ps1+\p2 & SU(3): 2\p1+2\ps1 & (1) & [\;\;] \\
{\bf  NR}&  SU(5):\p2 +\ps1&{\bf TDF} &&& SU(5) \\
(11) &  SU(6): \p3+s_2(\p1+\ps1) & \p1+\ps1 & SU(5):
  \p2+\ps2  &  & \\
  &  &  & \;\;\;\;+(s_2-1)(\p1+\ps1)  & (5,6) & \left( SU(3-s_2) \right) ^2 \\
{\bf  A}& SU(6):\p3& \p3 &&& SU(3)^2 \\
(13) & SU(7): \p3 + 2\ps1 &  \p3 & G_2: 2 \p1 & (60) &SU(2) \\ \hline  \hline
   & \mathbf{SO(2n+1)}\;\;
${\bf theories}$  &<\;>&${\bf Higgs flow}$&$entry$& 
   ${\bf SYM}$\\ \hline 
(14)&k\p1: 2n-3 \neq k\ < 2n-1 & \p1 & SO(2n): 
   (k-1)\p1 & (32)  & SO(2n+1-k)\\ 
{\bf  A}&(2n-3)\p1& \p1 & SO(2n): (2n-4)\p1 & (33) &  
   \left(SU(2)\right) ^2\\
(16) & SO(7): s_3\p1+S &\p1 &  SU(4): (s_3-1)\p2+\p1+\ps1 &(1,2,8)&  
   SU(4-s_3) \\
(17)& SO(7): (1+s_3)S & S & G_2: s_3\p1 &(60)&SU(4-s_3)\\
(18) & SO(7):S &S&&& G_2\\
(19) & SO(7): s_2\p1+2S&\p1 &  SU(4):(s_2-1)\p2+2(\p1+\ps1)& (1,2) & 
    SU(3-s_2) \\ 
(20) &SO(7): \p1 +3S &\p1 & SU(4):3(\p1+\ps1) &(1)& [\;\;] \\
(21) & SO(9): (1+s_3)\p1+S & \p1 &
      SO(8): s_3\p1 + S + S' &(36) & SU(4-s_3) \\
(22) & SO(9): \p1 + S & \p1 & SO(8):S + S' & (34) & G_2 \\
(23) & SO(9): S & S&&& SO(7)\\
(24) &SO(9): (s_3-1)\p1+2S &S& SO(7): \p1 + s_3S & $16,19,20$& 
      SU(4-s_3)\\
(25) &SO(9): 3S &S& SO(7): 2(\p1+S) &(19) & [\;\;] \\
(26) &SO(11): (1+s_3)\p1+S& \p1 &  SO(10):s_3\p1+S+S' & (42)& SU(4-s_3)\\
(27) &SO(11): \p1+S& \p1 & SO(10): S+S' &(44)&SU(4) \\
(28) & SO(11): S &S&&& SU(5) \\
(29) & SO(11): 2S &S& SU(5): \p2+\ps2+\p1+\ps1 &(5)& [\;\;]\\
(30) & SO(13): s_2\p1+S & \p1 & SO(12): (s_2-1)\p1+S+S' &(51,52)& 
     (SU(3-s_2))^2\\
(31) & SO(13): S & S &  SU(6): \p3 &(12)& \left( SU(3) \right)^2 \\ 
      \hline \hline 
& \mathbf{SO(2n)}\;\;  ${\bf theories}$  &<\;>&
${\bf Higgs flow}$&$entry$& ${\bf SYM}$\\ \hline 
(32)&k\p1:2n-4 \neq k < 2n-2& \p1 &
    SO(2n-1):(k-1)\p1& (14) &SO(2n-k) \\
{\bf A}&(2n-4)\p1 & \p1 & SO(2n-1): (2n-5)\p1 & (15) & 
    \left( SU(2)\right)^2\\ 
(34) & SO(8): \p1+S & \p1 & SO(7): S &(18) &G_2 \\
(35) & (2-s_2+s_3)\p1+s_2S & \p1 & 
    SO(7):(1-s_2+s_3)\p1+s_2S & (16,19) &SU(4-s_3)\\
(36) & SO(8): s_3\p1+S+S' & \p1 & SO(7): (s_3-1)\p1+2S & (17,19)&  SU(4-s_3) \\
(37) & SO(8): s_2\p1+2S +S' & \p1 &  SO(7): (s_2-1)\p1+3S &(17,20)&  
    SU(3-s_2) \\
(38) & SO(10): (2+s_3)\p1+S & \p1 & SO(9): (1+s_3)\p1+S & (21) & SU(4-s_3)\\
(39) & SO(10): 2\p1+S & \p1 & SO(9): \p1+S & (22) & G_2 \\
(40) & SO(10): \p1+S & \p1 & SO(9): S & (23) & SO(7) \\
{\bf NR}& SO(10):S  &{\bf TDF}&&& SO(10) \\ 
(42) & s_3\p1+s_2S+(2-s_2)S' & \p1 & SO(9):(s_3-1)\p1+2S&(24)&
     SU(4-s_3)\\
(43) & SO(10): 2S & 2S &&&   G_2 \\
(44) & SO(10): S + S' & S+S' & SU(5):\p1 +\ps1&(1)& SU(4)\\
(45) & \p1+(1+s_2)S+(2-s_2)S' & \p1 & SO(9): 3S & (25) &  [\;\;] \\
(46) & SO(10): 3S & 2S & G_2: 2\p1&  (60) & SU(2) \\
(47) & SO(10): 2S + S' & 2S & G_2: 2\p1 &  (16) & SU(2)\\
(48) & SO(12): s_5\p1+S & \p1 &  SO(11):(s_5-1)\p1+S &  26-28 & SU(6-s_5) \\
{\bf A} & SO(12): 2S & S & SU(6): \p2+\ps2 & (7) & (SU(2))^3 \\ 
(50) & SO(12): S & S & &  &SU(6)\\
(51) & \p1+s_2S+(2-s_2)S' & \p1 &  SO(11): 2S & (29) &  [\;\;] \\
(52) & SO(12): S+S' & S &   SU(6): \p3+\p1+\ps1 & (11) & 
    \left( SU(2) \right)^2\\
(53)& SO(14): s_3\p1+S &  \p1 &   SO(13): (s_3-1)\p1+S & (30,31) &  
    \left( SU(4-s_3) \right) ^2 \\
{\bf A}& SO(14): S  &S  &&& G_2 \times G_2 \\ \hline \hline 
& \mathbf{SP(2n)}\;\; ${\bf theories}$ & <\;> & 
    ${\bf Higgs flow}$ & $entry$ & ${\bf SYM}$ \\ \hline 
(55) & Sp(2n): 2k\p1: k \leq  n & 2\p1 & Sp(2(n-1)): 2(k-1)\p1 &   
    (1,55) & Sp(2(n-k))\\
(56) & Sp(2n): \p2+2\p1 & 2\p1 & Sp(2(n-1)): \p2+2\p1 & 
    (1,56) & [\;\;]\\
{\bf A} & Sp(2n): \p2  & \p2 &&&(SU(2))^n\\
(58) & Sp(4):2\p2 & \p2 & SO(4):\p1 &  (32) & SU(2) \\
(59) &  Sp(6): \p3+\p1 & \p3 & SU(3):\p1+\ps1& (1)  & SU(2)\\ \hline \hline
& ${\bf Exceptional groups}$ & <\;> & ${\bf Higgs flow}$ 
     & $entry$ & ${\bf SYM}$ \\ \hline 
(60) &G_2: s_3\p1 & \p1 & SU(3): (s_3-1)(\p1+\ps1) & (1)&SU(4-s_3)\\
(61) & F_4: 2\p1 & \p1 & SO(9): 2\p1+S & (21) &SU(3) \\
(62) & F_4: \p1 & \p1 &  SO(9): \p1 & (14) & SO(8) \\
(63) & E_6: s_2\p1+(2-s_2)\ps1  & \p1  & F_4: \p1 &(62)&  SO(8) \\
(64) &(1+s_2)\p1+(2-s_2)\ps1 & \p1 &   F_4: 2\p1 & (61) & SU(3)\\
(65) & E_6: \p1 & \p1 &&& F_4\\
(66) & E_7: 2\p1 & \p1 & E_6: \p1+\ps1 & (64) & SO(8) \\
(67) & E_7: \p1 &   \p1 &&&  E_6 \\
\end{longtable}
\end{small}

\section*{Appendix B \hspace*{0.25cm} Dimension of the classical moduli space}
\setcounter{equation}{0}
\renewcommand{\theequation}{B.\arabic{equation}}

This appendix is devoted to some more mathematical results. We define 
Zariski topology, which was mentioned in Section.~\ref{sec:mueq}. Next,
we make precise the notion of the dimension of the classical moduli space.
We explain a subtlety encountered for theories without classical
flat directions: $SO(10)$ with a spinor field and $SU(5)$ with
${\bf 10 + \bar{5}}$. The same subtlety arises also for theories which have
points in the moduli space with unbroken $SU(5)$ with ${\bf 10 + \bar{5}}$,
namely $SU(2 N+1)$ with $\Yasymm$ and $(2 N-3) \, \overline{\Yfund}$.

Let $V$ be an $n$ dimensional complex vector space, $V \cong {\bf C}^n$.  
In the Zariski topology,  a  closed subset $C \subseteq V$ 
is one that can be described as the set of zeros of a finite number of
polynomials,  i.e., $C = \{x \in {\bf C}^n | p_{\alpha}(x) = 0 ,
\alpha=1,...s \}  \equiv <~p_1,...,p_s>$. Open sets are those whose 
complements in $V$ are closed. The definition of Zariski 
closed sets satisfies  the axioms  of a topological space:
({\it i})~$V$ and $\emptyset$ are both open and closed,
({\it ii})~a finite union of closed sets is closed, and
({\it iii})~an arbitrary intersection of closed sets is closed.
Property ({\it iii}) follows from Hilbert's basis theorem, according to which
the zero set of an arbitrary set of polynomials agrees with the 
zero set of some {\em finite} set of polynomials. 

Any non-empty Zariski open subset $O \subseteq V$ is 
dense in $V$, i.e., the smallest closed set containing $O$ 
is $V$ itself. For this reason, a property 
satisfied by every point in a Zariski open subset
of a vector space $V$ is said to hold at  ``points at generic position''.
This makes precise the notion of ``photons in the bulk of the moduli
space'' introduced in Section~\ref{sec:mueq}. 

Another important property of non-empty Zariski open sets is that any two 
of them intersect non-trivially. The non-empty intersection 
is an open set, and therefore dense in $V$. 
An example of a Zariski closed set is the classical 
moduli space ${\cal M}_c \subseteq V$, $V$ the span of a basic 
set of gauge invariant polynomials  $p (\phi)$ on the elementary fields
$\phi$. ${\cal M}_c$ consists of the zeros of the polynomial
constraints satisfied by  the generators. The precise definition of the
classical moduli space follows. 

Having defined the Zariski topology we now discuss the classical moduli
space and its dimension. Consider the vector space $U$ of microscopic
fields $\phi^i$ in a gauge theory. Polynomials $p(\phi)$ on the chiral
matter fields $\phi^i$ form an algebra. There is a representation of the
gauge group $G_r$ on this algebra, namely
\begin{equation} 
\label{gact}
g \cdot p (\phi) = p (g^{-1} \phi), 
\end{equation}
which naturally extends to a representation of the complexification $G$ 
of $G_r$. If $G_r$ is the product of a compact, connected semi-simple 
Lie group with (possible) $U(1)$ factors, then the subalgebra of  
gauge invariant polynomials is finitely generated~\cite{gw}.
That means that a minimal set of basic gauge invariant polynomials 
$p_1(\phi),\cdots,p_n(\phi)$ can be found, in terms of which
any gauge invariant polynomial $p(\phi)$ can be written as 
\begin{equation}
\label{gen}
p(\phi) = \hat{p}(p_1(\phi),...,p_n(\phi)),
\end{equation}
where $\hat{p}(p_1,...,p_n)$ is a polynomial function. 
There is no unique choice for the basic invariants, 
but they can be chosen to be homogeneous, of degrees 
$d_1 \leq d_2 \leq \cdots \leq d_n$, and the sequence of 
degrees is uniquely determined by the $G$ representation~\cite{gw}.
In general, the basic invariants are constrained by a set
of equations of the form
\begin{equation}
\label{cons}
k_{\alpha}(p_1,...,p_n) = 0;
\end{equation} 
meaning that when evaluating $p_i = p_i(\phi)$ in the above equation 
we get zero. The set of basic invariants together with the relations
in Eq.~\ref{cons} is called classical moduli space, because their
points are in  one to one correspondence with D-flat
configurations~\cite{ProcSchwarz,BDFS,GA-NPB}. In some cases there are no
constraints among the $p_i'$s, the algebra of gauge invariant polynomials 
is freely generated by $p_1,..., p_n$.

We now consider the vector space $U$ of microscopic fields
$\phi^i$. The set $C_d \subseteq U \cong {\bf C}^n$
of points whose orbits under the action of the group $G$ have
dimension less than $d$ is closed. $C_d$ is closed because it is the
set of zeros of all polynomials in $\phi$ obtained by taking
the determinants of $d \times d$ submatrices of the $d_G \times n$ matrix
expressing the action of $G$ on $\phi$. The $d_G$ columns of this matrix
are $T_A \phi$, where $T_A$ are the generators of ${\rm Lie}\, G$,
$A=1,\ldots,d_G$. Taking $d$ to be equal to the maximum dimension of
an orbit we learn that all points with orbits of maximal dimension
form an open set, ${\cal O}_1'$, the complement of $C_{max}$.
Therefore a generic configuration of VEVs of $\phi$'s has orbits of
maximal dimension, or the maximal number of broken generators.

It is shown in Ref.~\cite{Elashvili} that 
${\cal O}_1'$ contains an open subset with the property that at every point
$\phi \in {\cal O}_1$, the subspace of unbroken generators is conjugate
to a fixed subalgebra ${\rm Lie}\, G_*$ of ${\rm Lie}\, G$.
The group $G_*$ is called the {\em stabilizer at general positions}.
Note that we have not yet commented on the possibility of breaking
$G$ to $G_*$ by a D-flat VEV\@. Whenever $G_*$ is
reductive\footnote{This means that every representation of $G_*$ can be
broken up into irreducible blocks. This is always true if Lie $G_*$ is
the sum of a semi-simple Lie algebra and (possible) $U(1)'s$.},
there exists a Zariski open, $G$ invariant set ${\cal O}_2\subset U$
containing closed orbits of $G$\@. This is equivalent to saying that
every point in the open set ${\cal O}_2$ is gauge related to a D-flat
point~\cite{ProcSchwarz,GA-NPB}. As ${\cal O}_1 \cap {\cal O}_2 \neq \emptyset$
and open we conclude that whenever $G_*$ is reductive there is
always a D-flat point that breaks $G$ to $G_*$\@. 
Moreover, since ${\cal O}_1 \cap {\cal O}_2$ is open, the set of D-flat
points breaking $G$ to $G_*$ is dense. We will label the minimal unbroken
subgroup at the D-flat points $G_P$. When $G_*$ is reductive, $G_P=G_*$.

There are only two physical (gauge anomaly free) theories 
for which $G_*$ fails to be reductive: $SU(2k+1)$ 
with an antisymmetric tensor and $2k-3$ antifundamentals, 
and $SO(10)$ with a spinor, see Table~\ref{fvt}. 
These are the only theories for which 
the $D$-flatness condition restricts the breaking of the gauge group 
to a subgroup $G_P$ of higher dimension than $G_*$. 
For all other theories $G_* = G_P$ and 
${\cal O}_1 \cap {\cal O}_2$ is  a dense open set of 
points with a closed $G$ orbit of maximum dimension, from which follows that
(Theorem 2 in Ref.~\cite{GA-NPB})
\begin{equation}
\label{dim}
    {\rm dim}\, {\cal M}_c = {\rm dim}\, U - {\rm dim}\, G + {\rm dim}\, G_P.
\end{equation}
Note that the r.h.s.\ of (\ref{dim}) equals ${\rm dim}\, U$ minus the maximal 
dimension of a $G$ orbit in $U$. Note also that for the theories in 
Table~\ref{fvt} eq.~(\ref{dim}) holds if we replace $G_P$ with 
$G_*$.

The dimension of ${\cal M}_c$  is much easier to compute using 
Eq.~\ref{dim} than its algebraic 
definition~\cite{gw}, which is a minimal number $r$ of gauge 
invariant polynomials $\chi^1(\phi),...,\chi^r(\phi)$ 
for which every polynomial in a basic set of invariants 
$p^1(\phi),...,p^k(\phi)$ satisfies an 
equation of the type
\begin{equation}
\label{algdim}
  (p^j)^t +(p^j)^{t-1}q_1(\chi^1,...,\chi^r) 
+ \cdots +  (p^j) q_{t-1}(\chi^1,...,\chi^r) 
+ q_t(\chi^1,...,\chi^r) = 0.
\end{equation}
The above equation makes precise the notion of ``number of 
basic invariants minus number of independent constraints.'' 

As an example, consider supersymmetric QCD with the number of colors
equal to the number of flavors and the usual choice of 
basic invariants $M^i_j = Q^{i\alpha}\tilde{Q}_{\alpha j}, B = \det Q , 
\tilde{B} = \det \tilde{Q}$, where $Q^{i \alpha}$ are the quarks and 
$\tilde{Q}_{j \beta}$ the anti-quarks. Here, $i,j$ are flavor
indices, $i,j=1,\ldots,N$, and $\alpha,\beta=1,\ldots,N$ are the
color ones.  A set of $\chi$'s is 
$\{M^i_j, \tilde{B}-B \}$. In fact 
\begin{equation} 
\tilde{B}^2 - (\tilde{B}-B)\tilde{B} - \det M = 0 \hspace{1cm} 
B^2 + (\tilde{B}-B)B  - \det M = 0, \\
\end{equation}
and it is easy to see that there is no smaller set of invariants 
satisfying~(\ref{algdim}) for supersymmetric QCD. 
This tells us that dim ${\cal M}_c = N^2 + 1$, which agrees
with~(\ref{dim}),  as $G_P$ is trivial. The definition and computation
of dimensions  when there is a tree level superpotential involves a number 
of additional subtleties~\cite{GA-NPB}. 

\begin{table}[tb] 
\caption[The only supersymmetric gauge theories with simple gauge 
group and non reductive stabilizer at general position $G_*$.]{\label{fvt}
\noindent The only supersymmetric gauge theories with simple gauge 
group and non reductive stabilizer at general position $G_*$. $G_P$  
is the unbroken subgroup at D-flat points where $G$ is maximally broken, 
and equals $G_*$ when $G_*$ is reductive.
$d_{{\cal M}_c}$ is the dimension of the moduli space. In the last column 
we give a basic set of gauge invariants, its number equals $d_{{\cal M}_c}$ 
because there are no constraints among them. ${\mathfrak{u}}(n)$ denotes 
the Lie algebra of a unipotent group of dimension $n$, which is 
the ``non-reductive piece'' of $G_*$.}  
\smallskip
\setlength{\tabcolsep}{1mm}
\renewcommand{\arraystretch}{1.5}
\newcolumntype{C}{>{$}c<{$}}
\begin{tabular}{|C|C|C|C|C|C|} \hline 
G & \rho & {\rm Lie}(G_*)&{\rm Lie}(G_P) & d_{{\cal M}_c} &
       {\rm Invariants}\\ \hline \hline
SO(10) & {\rm spin} & {\mathfrak{so}} (7) + {\mathfrak{u}} (8) &
   {\mathfrak{so}}(10) & 
0 & {\rm no\ invariants} \\ 
SU(2k+1) & \Yasymm + (2k-3) \overline{\Yfund} &
  {\mathfrak{su}}(2)+{\mathfrak{u}}(6)& 
{\mathfrak{su}}(5)
& (2k-3)(k-2) 
& A^{\alpha \beta} \bar Q _{\alpha}^i \bar Q _{\beta}^j\\
\hline
\end{tabular}
\end{table}

\end{document}